\documentclass{article}
\usepackage{arxiv}
\usepackage{multicol}
\usepackage[numbers,sort&compress]{natbib}
\usepackage[many]{tcolorbox}
\usepackage[utf8]{inputenc} % allow utf-8 input
\usepackage[T1]{fontenc}    % use 8-bit T1 fonts
\usepackage{url}            % simple URL typesetting
\usepackage{booktabs}       % professional-quality tables
\usepackage{amsfonts}       % blackboard math symbols
\usepackage{nicefrac}       % compact symbols for 1/2, etc.
\usepackage{microtype}      % microtypography
\usepackage{amsmath, amsthm}

\usepackage{color}
\usepackage{graphicx}       %Show Figs
\usepackage{float} %float figs
\usepackage{csquotes}
\usepackage{makecell}
\usepackage{hyperref}   
\usepackage{xcolor}
\usepackage{arxiv}
\usepackage{multicol}
\usepackage[capitalise,nameinlink,noabbrev]{cleveref}% hyperlinks
\hypersetup{               
    colorlinks=true,                
    breaklinks=true,                
    urlcolor= black,                
    linkcolor= red,                
    bookmarksopen=false,
    filecolor=black,
    citecolor=blue %blue, %if we want to enable/disable citation colors alone
}
\begin{document}
\title{Generative Models of Brain Dynamics}
% \title{Dynamical Systems and Machine Learning \\
% A Review and Perspectives for Neuroscience}
% \title{A tale of three worlds: AI, Neuroscience, and DST}
% \title{From Computational Models to Generative Models}
% \title{Machine learning and Dynamical Systems for neural generative models}
% \title{Brain dynamics and machine learning algorithms}

   \author{Mahta Ramezanian Panahi\thanks{Correspondings may be sent to mahtaa [at] gmail [dot] com}\\ \small{Mila - Quebec AI Institute}\\ \small{Université de Montréal, Montréal, Quebec, Canada}
      \And
      Germán Abrevaya \\\small{Departamento de Física, FCEyN, UBA}\\ \small{IFIBA, CONICET, Buenos Aires, Argentina}
      \And
      Jean-Christophe Gagnon-Audet\\ \small{Mila - Quebec AI Institute}\\ \small{Université de Montréal,
      Montréal, Quebec, Canada}
      \And
      Vikram Voleti\\ \small{Mila - Quebec AI Institute}\\ \small{Université de Montréal,
      Montréal, Quebec, Canada}
      \And
      Irina Rish %\thanks{Senior co-authors contributed equally.},
      \\ \small{Mila - Quebec AI Institute}\\ \small{Université de Montréal,
      Montréal, Quebec, Canada}
      \And
      Guillaume Dumas\\ \small{Mila - Quebec AI Institute}\\ \small{CHU Sainte-Justine Research Center, Department of Psychiatry,}  \\ \small{Université de Montréal,
      Montréal, Quebec, Canada}
  }
\date{}

\maketitle
% \doublespacing
\begin{abstract}
%%%% CONTEX: %%%%%
The principled design and discovery of biologically- and physically-informed models of neuronal dynamics has been advancing since the mid-twentieth century. Recent developments in artificial intelligence (AI) have accelerated this progress. 
%%%%%% THESIS STATEMENT %%%%%%%
This review article gives a high-level overview of the approaches across different scales of organization and levels of abstraction.
%%%%%% Descriptive sentence %%%%%%
The studies covered in this paper include fundamental models in computational neuroscience, nonlinear dynamics, data-driven methods, 
% NOTE: The word phenomenological is deliberately avoided because it might be confusing in an out-of-context abstract
as well as emergent practices. While not all of these models span the intersection of neuroscience, AI, and system dynamics, all of them do or can work in tandem as generative models, which, as we argue, provide superior properties for the analysis of neuroscientific data. 
%%%%%%%%%%% SUMMARY %%%%%%%%%%
We discuss the limitations and unique dynamical traits of brain data and the complementary need for hypothesis- and data-driven modeling.
%%%%%%%%%% CONCLUSION %%%%%%%%
By way of conclusion, we present several hybrid generative models from recent literature in scientific machine learning, which can be efficiently deployed to yield interpretable models of neural dynamics. 

\end{abstract}
\vspace{1cm}
\begin{multicols}{2}
\section*{Introduction}
\label{sec:Intro}
%*******What and why computational models********
\begin{displayquote}
 \emph{“What I cannot create I do not understand.”} --- Richard Feynman
\end{displayquote}

\begin{figure*}[!hbp]
    \centering
    \includegraphics[width=0.7\linewidth]{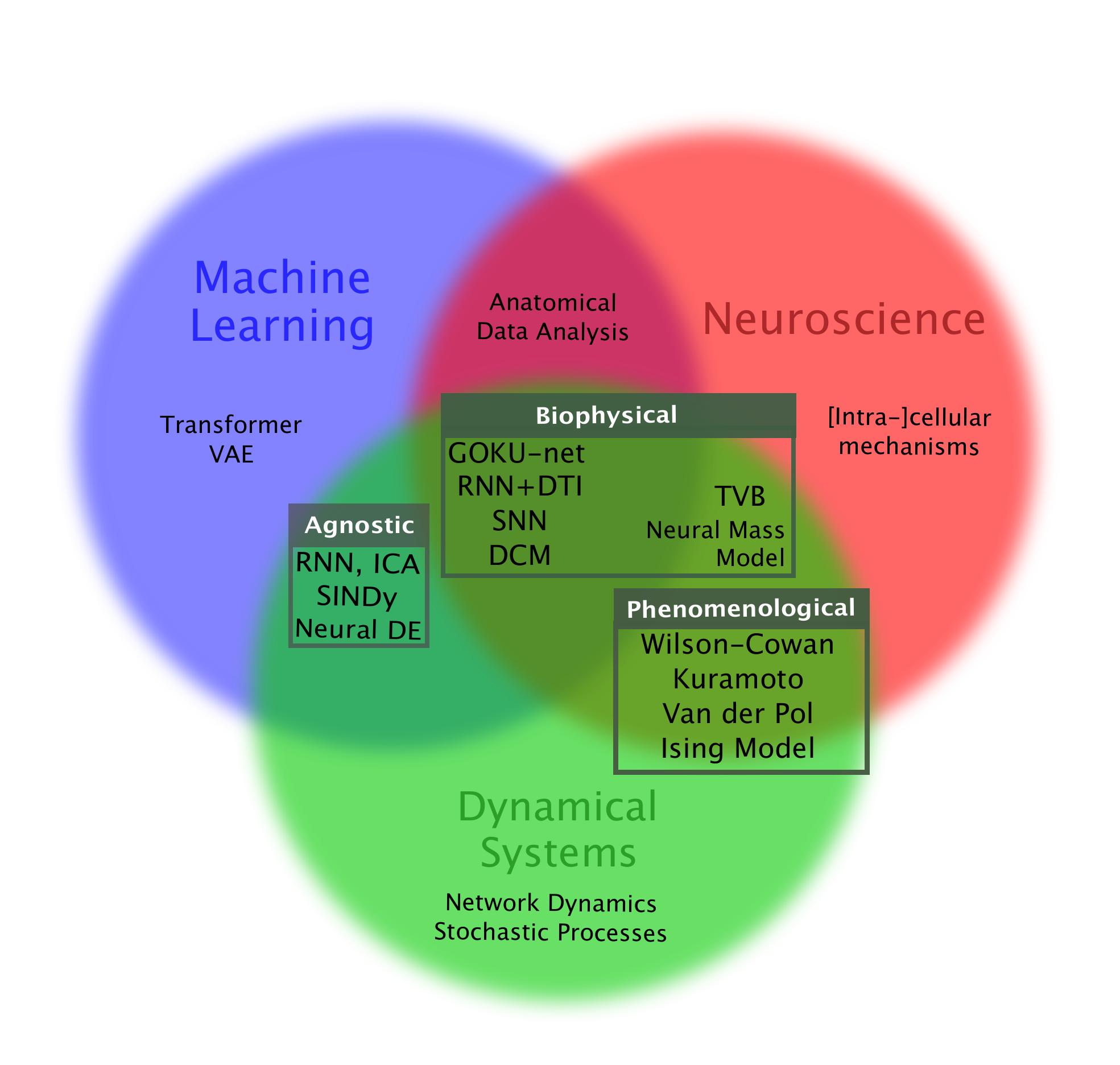}
    \caption{\textit{Venn diagram of the generative models of interest. 
    Based on the abstraction and assumption, methods might belong to one or more of the three worlds of machine learning, neuroscience, and dynamical systems.
    This review is structured into three main categories that are in fact, intersections of these fields: biophysical (Section \ref{sec:bio}), phenomenological (Section \ref{sec:phenomenological}), and agnostic modeling (Section \ref{sec:agnostic_computational}).
 Tools developed independently in each of these fields can be combined to overcome the limitation of data.}}
    \label{fig:scope_venn}
\end{figure*}

The explosion of novel data acquisition and computation methods has motivated neuroscientists to tailor these tools for ad hoc problems. While attempts at pattern detection in enormous datasets are commonplace in the literature– representing a logical first step in applying learning algorithms to complex data—such efforts provide little insight into the observed mechanisms and emission properties. As the above quote from R. Feynman suggests, such methods are \textit{understanding} the brain. 
The importance of developing interpretable algorithms for biological data– beyond the standard “black-box” models of conventional machine learning–is underscored by the pressing need for superior explainability seen in medical and health-related research. To this end, formal modeling (the practice of expressing some dependent variable unequivocally in terms of some other set of independent variables \cite{wills2012adequacy}) is the only way for transparent and reproducible theories \cite{guest2020computational}. In the present review, we propose that a class of architectures known as generative models constitute an emergent set of tools with superior properties for reconstructing segregated and whole-brain dynamics. A generative model may consist of, for example, a set of equations that determine the evolution of the signals from a human patient based on system parameters. In general, generative models have the benefit over black-box models containing inference mechanisms rather than simple predictive capacity.

\paragraph{Why prefer inference over prediction?} Put simply: the goal of science is to leverage prior knowledge, not merely to forecast the future (a task well suited to engineering problems), but to answer “why,” questions, and to facilitate the discovery of mechanisms and principles of operation. \citet{bzdok2019exploration} discuss why inference should be prioritized over prediction for building a reproducible and expandable body of knowledge.
We argue that this priority should be especially respected for clinical neuroscience. 

It is important to note that modeling is, and should be, beyond prediction \cite{epstein2008model}. Not only does explicit modeling allow for explanation (which is the main point of science), but it also directs experiments and allows for the generation of  new scientific questions. 

In this paper, we demonstrate why focusing on the multi-scale dynamics of the brain is essential for biologically plausible and explainable results. For this goal, we review a large spectrum of computational models for reconstructing neural dynamics developed by diverse scientific fields, such as biological neuroscience (biological models), physics, and applied mathematics (phenomenological models), as well as statistics and computer science (data-driven models). On this path, it is crucial to consider the uniqueness of neural dynamics and the shortcomings of data collection.  
Neural dynamics are different from other forms of physical time series. In general, neural ensembles diverge from many canonical examples of dynamical systems in the following ways: 

\paragraph{Neural dynamics is different.}

A neural ensemble is distinctive from the general notion of the dynamical system:
\begin{itemize}
    \item Unlike chemical oscillations and power grids, the nervous system is a product of biological evolution, which makes it special regarding complexity and organization.
    \item Like many biophysical systems, it is highly dissipative and functions in non-equilibrium regimes (at least while working as a living organ).
    \item Although the brain exhibits continual neuromodulation, the anatomical structure of the brain is encoded in the genome, hence it is essentially determined \cite{rabinovich2006dynamical}.  
    \item There are meaningful similarities in brain activity across species. This is especially good news because, unlike humans, neural properties  of less-complicated species are well-characterized \cite{white1986structure}. 
\end{itemize}
These characteristics help narrow down the search for useful models.

% \\On a more controversial side, some levels of autonomy for neural populations and even single neurons can be assumed based on the self-regulation of the subjective component. This assumption makes the first-order approximation of some dynamical behavior reducible to a few parameters. Finally, neuroscientists widely take advantage of a phenomenological approach which is useful in deriving dynamics across different neural architecture with similar emergent functions.

\paragraph{Neural data is different.}
  Neural recordings - especially of human subjects - are noisy and often scarce.  Due to requirements of medical certification, cost of imaging assays, and challenges with recruitment, acquiring these datasets can be both expensive and time-consuming. Moreover, such data can be difficult to wrangle and contains inconsistent noise –not only across participants, but quite often for a single participant at different times (e.g., artifacts, skin condition, and time resolution in the case of EEG).

\subsection*{Overview of generative models}
\begin{figure*}[!hbp]
    \centering
    \includegraphics[width=0.7\linewidth]{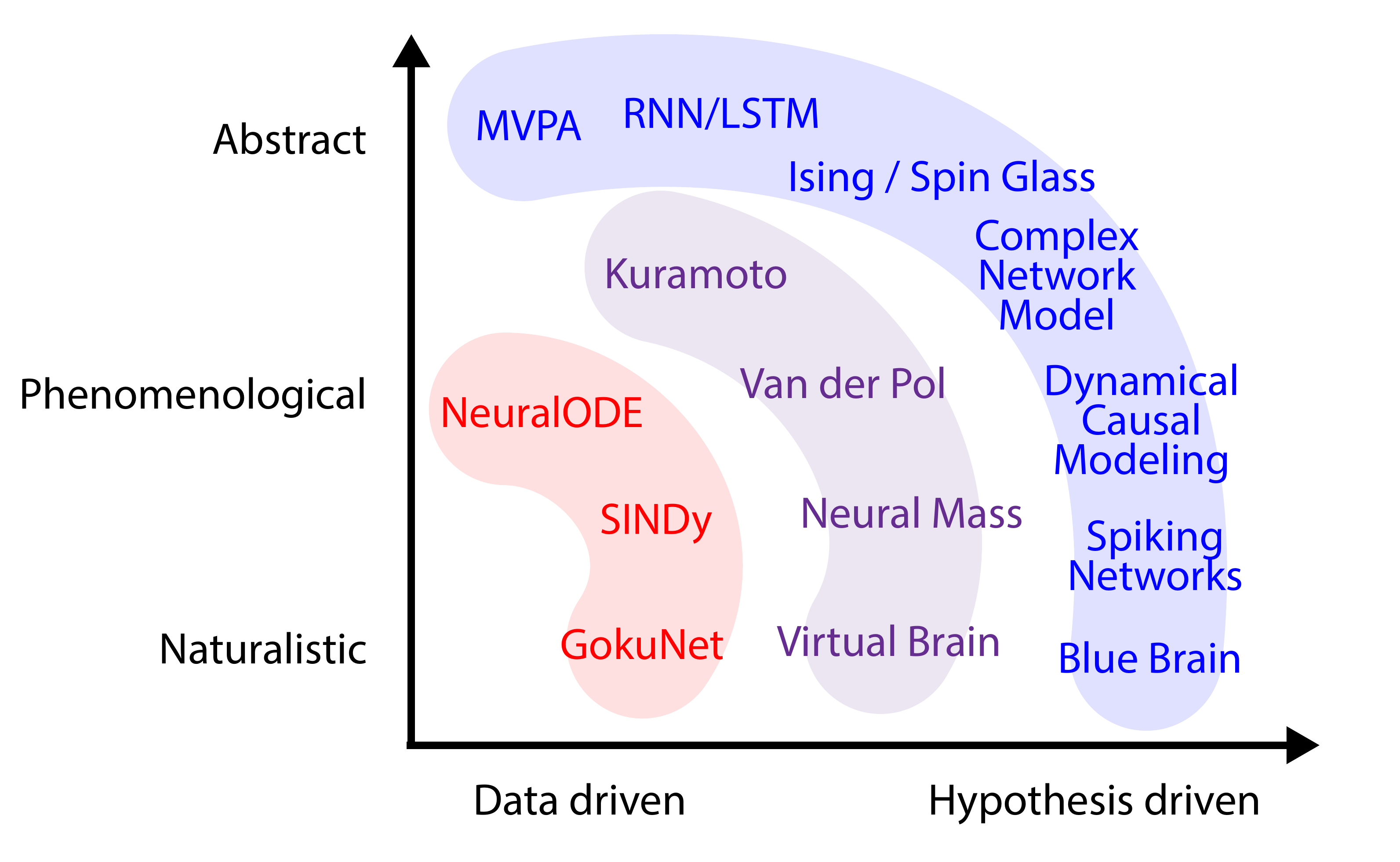}
    \caption{\textit{Overview of generative models: well-developed models (blue), partially-explored approaches (purple), and modern pathways with little or no literature on neural data (red)}.}
    \label{fig:methods_landscape}
\end{figure*}

Current generative models fall into three main categories as shown in Figure \ref{fig:scope_venn} with respect to their modeling assumption and objective:
\begin{enumerate}
    % TODO explain the distinction
    \item \textbf{Biophysical models:} 
    Biophysical models are \textit{realistic} models which encapsulate biological assumptions and constraints. Due to large number of components and the empirical complexity of the systems modeled,  examples of biophysical models run the gamut, from very small, with a high degree of realism (e.g., Hodgkin and Huxley’s model of squid giant axon), to large scale (e.g., \citet{Izhikevich2008large} model of whole cortex). Due to computational limitations, large-scale models are often accompanied by increasing levels of simplification. Blue Brain Project \cite{markram2006blue} is an example of this type of modeling.
    
    \item \textbf{Phenomenological models:}
        Analogies and behavioral similarities between neural populations and established physical models open the possibility of using well-developed tools in Statistical Physics and Complex Systems for brain simulations. In such models, some priors of the dynamics are given but not by realistic biological assumptions. A famous example is the model of Kuramoto oscillators \cite{Bahri2020} in which the goal is to find the parameters that best reconstruct the behavior of the system. These parameters describe the property of the phenomenon (e.g., the strength of the synchrony), although they do not directly express the fabric of the organism.
    
    \item \textbf{Agnostic computational:}
    Data-driven methods that, given a ``sufficient'' amount of data, can \textit{learn} reconstruct the behavior 
    % (e.g., distribution)
    with little prior knowledge. Examples of such approaches are some self-supervised methods such as latent ODEs \cite{chen2018neural}. The term ``sufficient'' expresses the main limitation of these approaches. Such approaches often need unrealistically large datasets and come with intrinsic biases. In addition, the representation that these models provide can be analytically far from the physics of the system or the phenomenon.

Figure \ref{fig:methods_landscape} shows an overview of various generative models and the presence in the literature up to this date.

\end{enumerate}
\paragraph{Key Contributions:} 
The objective is to bridge a gap in the literature of computational neuroscience, dynamical systems, and AI and to review the usability of proposed generative models concerning the limitation of data, the objective of the study and the problem definition, prior knowledge of the system, and sets of assumptions (see Figure \ref{fig:methods_landscape}).

We argue that each set of approaches presented here can facilitate hybrid solutions by borrowing essential ideas from other domains (e.g., computer vision and natural language processing) to model brain recording. 
\section{Biophysical Models}
\label{sec:bio}
%%%%%%%%%%%%%%%%%%%%%
% MOTIVATION
%%%%%%%%%%%%%%%%%%%%%

Understanding how cognition ``emerges'' from complex biophysical processes has been one of the main objectives of computational neuroscience.
Although inferring high-level cognitive tasks from biological processes is not easily achieved,  different biophysical simulations provide some "explanation" of how neural information relates to behavior. Those attempts are motivated by the need for interpretable and biologically-detailed models. 

\begin{figure*}[!hbp]
    \centering
    \includegraphics[width=0.9\linewidth]{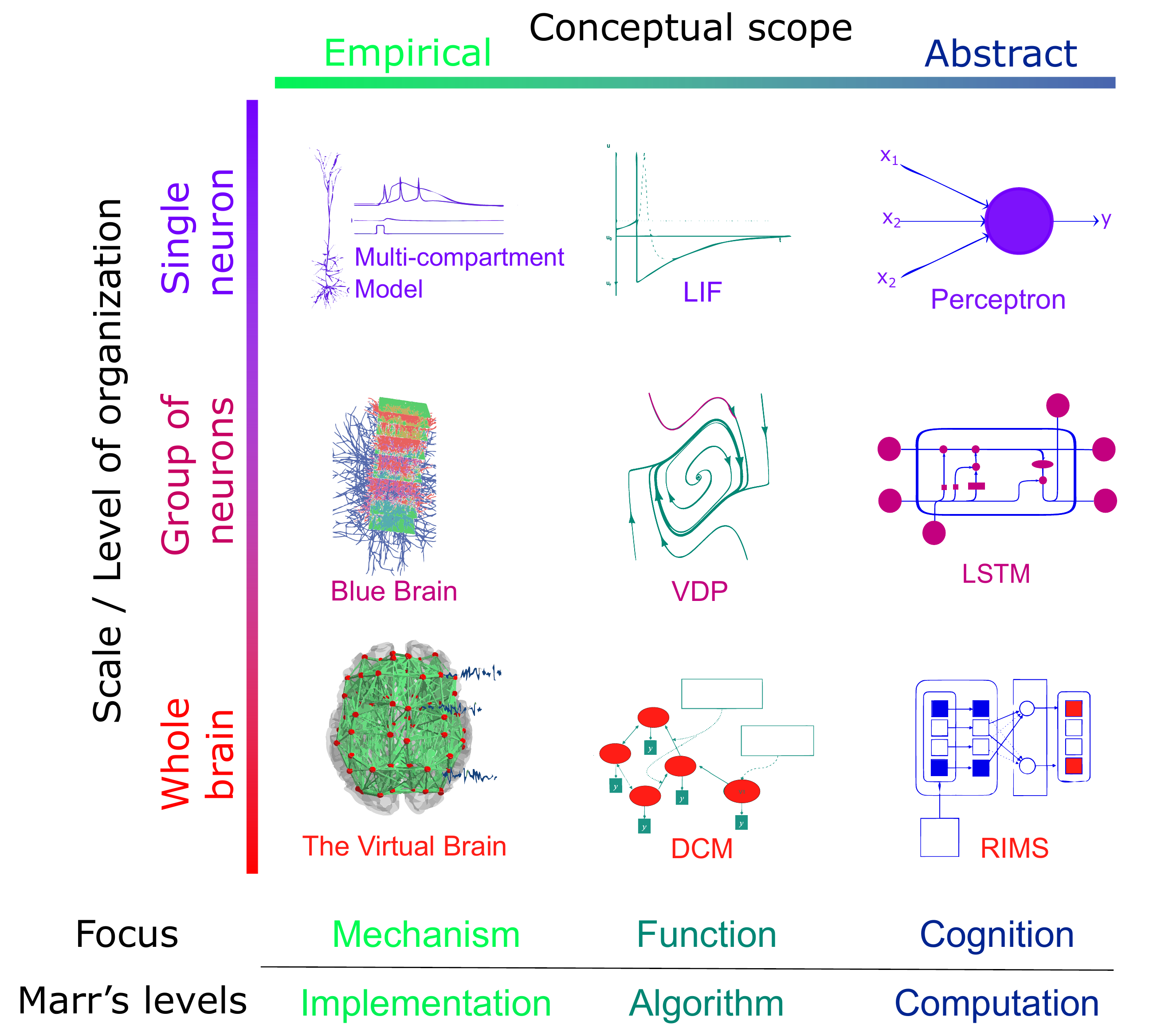}
    \caption{\textit{Instances of modeling across different levels of the organization and problem dimension. The \textit{conceptual scope} is an indicator of biophysical details incorporated in the model. It determines how the focus of the model is directed toward mechanistic reality or the behavioral output. It is also an indicator of where a given model sits on the Marr's level.}}
    \label{fig:scales_levels}
\end{figure*}

While there is as yet no ``unifying theory of Neuroscience'', biological neuronal models are being developed in different scales and with different degrees of abstraction (see Figure \ref{fig:scales_levels}). These models are usually grouped into two main categories:the first represents a ``bottom-up'' approach, which emphasizes biophysical details for fine-scale simulation and expects the emergence\footnote{\label{fn:emergence} Emergence is the manifestation of collective behavior that cannot be deduced from the sum of the behavior of the parts   \cite{johnson2002emergence, Krakauer2017Neuroscience}.}.
% \footnote{\label{fn:emergence}Emergence as defined in \cite{johnson2002emergence}} %%TODO elaborate on footnote
 An example of this approach is the Blue Brain project \cite{markram2006blue}.
 
 Conversely, ``top-down'' schemes focus on explicit high-level functions and designing frameworks based on some targeted behavior. Each of the two approaches works with a different knowledge domain and has its own pitfalls. The top-down approach can incorporate behavioral insights without concerning itself with hard-to-code biological details to generate high-level observed behavior. Models of this kind do not provide low-level explanations and are prone to biases related to data collection \cite{srivastava2020robustness}. The bottom-up perspective, on the other hand, benefits from a customized level of biophysical insight. At the same time, its description is not generalizable to behavior, and it can be difficult to scale (thanks to unknown priors and numerous parallel mechanisms). Also, the reductionist approach to complex systems (e.g., the brain) is subject to substantial criticism. In particular, while a reductionist approach can help to examine causality, it is not enough for understanding how the brain maps onto the behavior \cite{anderson1972more, Krakauer2017Neuroscience}.
%%%%%%%%%%%%%%%%%%%%%
% ORGANIZATION
%%%%%%%%%%%%%%%%%%%%%

In this section, we review brain models across different scales that are faithful to biological constraints. We focus primarily on the first column from the left in Figure \ref{fig:scales_levels}, starting from the realistic models with mesoscopic details to more coarse-grained frameworks.

\subsection{Modeling at the synaptic level}
The smallest interacting blocks of the nervous system are proteins \cite{Heuvel2019multiscale}. Genetic expression maps and atlases are useful for discovering the functions of these blocks in the neural circuit \cite{mazziotta2000probabilistic}.
However, these maps are not uniformly expressed in the brain \cite{lein2007genome}. While the expression maps of those proteins continue to unfold \cite{hawrylycz2012anatomically}, combined with connectivity data, they can help quantify dynamics. These maps link the spatial distribution of gene expression patterns, and neural couplings \cite{richiardi2015correlated} as well as other large-scale dynamics, e.g., dynamic connectivity as a dependent of neurogenetic profile \cite{diez2018neurogenetic}.

A notable effort in this regard is the Allen Brain Atlas \cite{jones2009allen} in which genomic data of mice, humans, and non-human primates \cite{hawrylycz2014allen} have been collected and mapped for understanding structural and functional architecture of the brain \cite{gilbert2018allen}. While genomic data by itself is valuable for mapping out connectivity in different cell types, a fifth division of AA, \textit{Allen Institute for Neural Dynamics} was recently announced, with the aim of studying the link between neural circuits of laboratory mice and behaviors related to foraging \cite{allen2021Anouncing}.

On a slightly larger scale, a considerable amount of work concerns the relationship between cellular and intracellular events and neural dynamics. Intracellular events and interactions models could generate accurate responses on small \cite{dougherty2005computational} and large scales \cite{whittington1995synchronized}. Some of these models laid the foundation of computational neuroscience and are reviewed in Subsection \ref{subsection:bio_history}. 
In what follows, we start with neurocomputational models at the mesoscale level (realistic models of small groups of neurons, i.e., the top-left corner of Figure \ref{fig:scales_levels}), after which we move on towards macro-scale levels with different degrees of abstraction.

\subsection{Basic biophysics of neurons: A quick history}
\label{subsection:bio_history}
Zooming out from the intra-neuron synaptic level, inter-neuron communication emerge as a principal determinant of the dynamics. Information transmission is mainly based on the emission of action potentials. The mechanism of this flow of ions was first explained by the influential Hodgkin-Huxley equations and corresponding circuits. The electrical current of the equivalent circuit is described by four differential equations that incorporate membrane capacity and the gating variables of the channels \cite{hodgkin1952quantitative}. While the Hodgkin-Huxley model agrees with a wide range of experiments \cite{traub1991model, patlak1985slow} and continues to be a reference for models of ion channels, it needs to be simplified in order to be expandable to the models of the neuronal population. The main difficulty with the Hodgkin-Huxley model is that it requires solving a system of differential equations for each of the gating parameters of each of the single ion channels of a cell while there are more than 300 types of ion channels discovered as of today \cite{gabashvili2007ion}. Various relaxing assumptions have been proposed, one of which is to dismiss the time dependence of membrane conductance and the dynamics of the action potential by simply assuming the firing happens when the electrical input accumulated at the membrane exceeds a threshold \cite{abbott1990model}. The latter model is known as integrate-and-fire \cite{stein1967frequency}, and it comes in different flavors depending on the form of nonlinearity assumed for the dynamics of leaky or refractory synapses \cite{michaels2016neural}.
% Note that although action potentials are important low-level modulators of population-level activity, they do not tell the whole story of cellular mechanisms. Alternately, \textit{neuromodulations} are also processes that influence the biophysics of neurons, such as gating parameters. Although their corresponding computations at a whole-brain scale is to be figured out, a recent study bridge these single-cell mechanisms to some cortical population gains \cite{Shine2021computational}. On another note, it is interesting to see how cross-scale studies are helpful in the testing theory of the smaller scales, which are hard to validate empirically.

In order to model interesting dynamics of various ion channels, a model of compartments of dendrites, called the multicompartment model, can be employed. An exclusive review article by Herz et al. \cite{herz2006modeling} categorizes compartmental models into five groups based on the level of balance and details involved from Hodgkin-Huxley description to black-box.

While the research on hyper-realistic modeling of many neurons continues, other frameworks focus on simulating the biophysics of the population of neurons. 
In the Subsection \ref{subsection:Population_bio}, we pause on the state of large-scale synaptic simulations to show how a change in computational paradigm helps in overcoming some of the limitations inherent in these models.
Models of Neural mass, Wilson-Cowan, and dynamical causation are examples of such alternatives (see Subsections \ref{subsection:Neural_mass_mean_field}, \ref{subsubsection:Wilson-Cowan}, and \ref{subsubsection:DcM} respectively). 
% NEST is open-source and different research groups are involved in development od \cite{eppler2009pynest}
% GENESIS \cite{bower2012book} and NEURON \cite{carnevale2006neuron} are two of the traditional toolboxes for this type of morphological reconstruction.

\subsection{Population-level models}
\label{subsection:Population_bio}
% Despite the difficulties and intractability of incorporating synaptic-level details for reconstructing collective behaviour, significant studies focus on such hyper-realistic approaches.

\citet{Izhikevich2008large} describe the first attempt in reconstructing the whole cortex. Their simulation includes a microcircuitry of 22 basic types of neurons with simplified dendrite trees and fewer synapses. The underlying structural data based on the geometry of the white matter is drawn from diffusion tensor imaging (DTI) \cite{honey2009predicting} of the human brain. The microcircuitry of the six-layered neocortex was reconstructed based on cats' visual cortex. The spiking dynamic employed in this model 
comes from \cite{izhikevich2003simple} and it is a simplification of the Hodgkin-Huxley model as it outputs the firing rates instead of currents. On a larger scale, some subcortical dynamics
(e.g., dopaminergic rewarding from the brainstem)
are also implemented.

The significance of this simulation compared to preceding efforts is its inclusion of all cortical regions and some of their interplays in the form of cortico-cortical connections. The researchers also considered synaptic plasticity a significant factor in studying developmental changes such as learning. The model demonstrates several emergent phenomena such as self-sustained spontaneous activity, chaotic dynamics, and avalanches, alongside delta, alpha, and beta waves, and other heterogeneous oscillatory activities similar to those in the human brain. 
% By contrasting with BOLD images passed through 0.1 - 0.01 Hz filter, the authors have found similar anti-correlated functional connectivities for the slow time scales (minutes). Also, the distribution of firing rates resembles the ones recorded \textit{in-vivo} for different types of neurons.
%  Further validation and development of metrics for measuring the similarity of the model and \textit{in-vivo} recordings are needed. The authors introduce two ways to look at the model's extreme sensitivity to the initial condition: $1)$ The spiking activity should be studied \textit{statistically} and the study of single-neuron should be disregarded because of the chaotic nature of the system. $2)$ There is a grain of truth in the highly-sensitive model as this sensitivity manifests itself in single-neuron stimulation effect on behavior in rats (\cite{houweling2008behavioural}) and the experimental evidence for microstimulation of a single tactile afferent in humans (\cite{vallbo1984microstimulation}).

Complexity aside, the model has its shortcomings, including extreme sensitivity to the initial condition. To address this, the authors suggest studying the population behavior instead of single-cell simulations. Despite all the limits, \cite{Izhikevich2008large} is the first benchmark of whole-cortex modeling and the foundation of future detailed projects such as Blue Brain project \cite{markram2006blue} and MindScope \cite{koch2014project}.
% Although, few observations in rats (\cite{houweling2008behavioural}) and humans (\cite{vallbo1984microstimulation}) verify the existence of chaotic behavior to some degree, but generally, this hyper dependency on the individual state of single neurons is in not consistent with reliable functionality. 
% The authors also expected that the model could be used as a prediction tool for practically-complicated experiments since an \textit{in-silico} environment provides a reproducible and noiseless environment. This feature opens new doors to studying the effect of changes in neural parameters for a variant set of neuronal pathologies and is particularly interesting when new dynamics -e.g., for sleep state-  are introduced into the system.

Following Izhikevich and Edelman \cite{Izhikevich2008large}, the Blue Brain Project \cite{markram2006blue} was founded in 2005 as a biological simulation of synapses and neurons of the neocortical microcircuitry. The ambitious goal was to extend this effort to a whole-brain level and build ``the brain in a box''. The initial simulated subject was only a $2 mm$ tall and $210 \mu m $ in radius fragment of the somatosensory cortex of a juvenile rat ($\sim100,000$ neurons). The efforts for further expansions to larger scales, i.e., mouse whole-brain and human-whole brain, are far-fetched by many critics \cite{abbott2020documentary}.

Far from the initial promise of ``{understanding}'' of the brain, the Blue Brain Project  is still far from incorporating  the full map of connections (also known as connectome \cite{HORN2014142}) in the mouse brain, which is still an order of magnitude smaller than the human brain \cite{fregnac2014neuroscience}. That being said, acquiring the connectomic map does not necessarily result in a better understanding of function. Note that while the connectomic structure of the roundworm Caenorhabditis elegans nervous system has been entirely constructed since 1986 \cite{white1986structure}, research is still unable to explain the behavior of the network, e.g., predicting stimuli based on excitation \cite{koch2012neuroscience}.
Finally, strong concerns regarding the validity of the experiments rise from the fact that the simulation still does not account for 
% the blood vessels and 
the glial cells. Glial cells constitute 90\% of the brain cells. They have distinctive mechanisms as they do not output electric impulses \cite{fields2014glial} but are responsible for inactivating and discharging products of neuronal activities which influence the synaptic properties \cite{Hamberger1971Glial} and consequently learning and cognitive processes \cite{fields2014glial}. This point of incompleteness sheds extra doubts on the achievability of \textit{brain in silico} from the Human Brain Project.  

The above critiques have been calling for a revision of the objectives of the Blue Brain project with more transparency. Hence, new strategies such as the division of Allen Institute, MindScope \cite{Hawrylycz2016Inferring}, and the Human Brain Project \cite{Amunt2016HBP} aim for adaptive granularity, more focused research on human data, and pooling of resources through cloud-based collaboration and open science \cite{fecher2014open}. Alternatively, smaller teams developed less resource-intensive simulation tools such as Brian \cite{stimberg2019brian} and NEST \cite{gewaltig2007nest}.

There are several readily-available simulators of large networks of spiking neurons to reconstruct many-neuron biophysics.
Brian \cite{stimberg2019brian} is a Python package for defining a customized spiking network. The package can automatically generate the code for simulating a computationally-optimal language (e.g., C++, Python, or Cython). With GPUs available, it can also enable parallelism for faster execution.
Brian is more focused on single-compartment models while GENESIS \cite{bower2012book} and NEURON \cite{carnevale2006neuron} center around multicompartment cells.

NEST is another popular package for building ad hoc models of spiking neurons with adjusted parameters. These parameters include the spiking rules (such as IF, Hodgkin-Huxley AdEx), networks (topological or random neural networks), synaptic dynamics (plasticity expressions, neuromodulation) \cite{gewaltig2007nest}. 

While working with mid-level packages, Technical limitations and the objective of the study should be considered. These include computational efficiency and the code generation pipeline.
Interested researchers are encouraged to refer to the review by Blundell et al. \cite{blundell2018code} to learn more about the guidelines and proposed solutions.

The steep price of high-resolution computation and the remoteness from high-level cognition can be levitated by replacing detailed dynamics of single neurons with the collective equations of the population. This dimensionality-reduction strategy is the essence of the neural mass models \cite{david2003neural},  spiking neural network \cite{vreeken2003spiking}, and dynamical causal modeling \cite{FRISTON2003DCM}.

\subsubsection{Neural mass models}
\label{subsection:Neural_mass_mean_field}
Staying faithful to the biophysical truth of the system can happen at scales larger than a few cells. 
% Mean-fields and neural masses are parsimonious estimations of the state of neural ensembles based on a functional differential equation (DE) of the probability density dynamics of a neuronal population.
In other words, by reducing the degrees of freedom, one can reduce a massive collection of individual integrate-and-fire equations (mentioned in Subsection \ref{subsection:bio_history}) to a functional DE of the probabilistic evolution of the whole population known as Fokker-Planck DE.
However, since Fokker-Planck equations are generally high-dimensional and intractable, a complimentary formalism, known as the mean-field approximation, is proposed for finessing the system \cite{deco2008dynamic}.

In statistical Physics, the mean-field approximation is a conventional way of lessening dimensions of a many-body problem by averaging over the degrees of freedom. A well-known classic example is the problem of finding collective parameters (such as pressure or temperature) of bulk of gas with known microscopic parameters (such as velocity or mass of the particles) by the Boltzmann distribution.
The analogy of the classic gas shows the gist of the neural mass model: the temperature is an emergent phenomenon of the gas \textit{ensemble}. Although higher temperatures correspond to higher \textit{average} velocity of the particles, one needs a computational bridge to map microscopic parameters to the macroscopic one(s). To be clear, remember that each particle has many relevant attributes (e.g., velocity, mass, and the interaction force relative to other particles). Each attribute denotes one dimension in the phase space. One can immediately see how this problem can become computationally impossible even for $1 cm^3$ of gas with $\sim 10^{19}$ molecules.

The current state of thermodynamics accurately describes the macroscopic behavior of gas, so why not use this approximation to the many-body problems of neuronal populations? The analogous problem for a neural mass model can be described with the single-neuron activity and membrane potential as the microscopic parameter and the state of the neural ensemble in phase space as the macroscopic parameter. The computational bridge is based on Fokker-Planck equations for separate ensembles.

Neural mass models can be used both for understanding the basic principles of neural dynamics and building generative models \cite{friston2008mean}. They can also be generalized to neural fields with wave equations of the states in phase space \cite{Coombes2005waves} as well as other interesting dynamical patterns \cite{coombes2014neural}. Moreover, these models are applicable across different scales and levels of granularity from subpopulations to the brain as a whole. This generalizability makes them a good candidate for analysis on different levels of granularity, ranging from modeling the average firing rate to decision-making and seizure-related phase transitions. The interested readers are encouraged to refer to the review in \cite{deco2008dynamic} to see how neural mass models can provide a unifying framework to link single-neuron activity to emergent properties of the cortex. Neural mass and field models build the foundation for many of the large-scale \textit{in-silico} brain simulations \cite{coombes2019next} and have been deployed in many of the recent computational environments \cite{jirsa2010towards, Ito2007Dynamics}. Note that the neural mass model can show inconsistency in the limits of synchrony and require complementary adjustments for systems with rich dynamics \cite{Deschle2020validity} by mixing with other models of neural dynamics such as Wilson-Cowan \cite{wilson1972excitatory} as in \cite{coombes2019next}.

% usually are based on systems of nonlinear ordinary differential equations. Wilson-Cowan is a notable instance in which the interaction of inhibitory and excitatory populations is described.
\subsubsection{Wilson-Cowan} \label{subsubsection:Wilson-Cowan}
Wilson-Cowan is a large-scale model of collective activity of neural population based on mean-field approximation (see Subsection \ref{subsection:Neural_mass_mean_field}). 
Seemingly the most influential model in computational neuroscience after Hodgkin-Huxley~\cite{hodgkin1952quantitative}
is Wilson-Cowan~\cite{wilson1972excitatory} with presently over 3000 mentions in the literature.

The significance of this work in comparison to its proceedings (e.g., in \cite{anninos1970dynamics, beurle1956properties}) is more than a formal introduction of tools from dynamical systems in neuroscience. This model acknowledges the diversity of synapses by integrating distinct inhibitory and excitatory subpopulations. Consequently, the system is described by two state variables instead of one.
Moreover, the model accounts for Dale's principle \cite{eccles1954cholinergic} for a more realistic portrayal. That is to say each neuron is considered purely inhibitory or excitatory.
The four theorems proved in the seminal paper \cite{wilson1972excitatory} conclude the existence of oscillations as a response to a specific class of stimulus configuration and the exhibition of simple hysteresis for other classes of stimulus.

Wilson-Cowan model lays the foundation for many of the major theoretical advances. Examples of the derivative studies include energy function optimization for formulating associative memory \cite{Hopfield1982neural}, artificial neural networks as a special case with binary spiking neurons \cite{hinton1983optimal}, pattern formation \cite{amari1977dynamics}, brain wave propagation \cite{roberts2019metastable}, movement preparation \cite{erlhagen2002dynamic}, and Dynamic Causal Modeling \cite{Sadeghi20DCM}. Other studies also demonstrate the possibility of diverse nonlinear behavior of networks of Wilson-Cowan oscillators \cite{Wilson2019Hyperchaos, MacLaurin2018mean}. More detailed extensions are on the way. For example, second-order approximations \cite{el2009master} and simulation of intrinsic structures such as spiking-frequency adaptation or depressing synapses~\cite{chen2018attractors}).
For a comprehensive list of continuations, see \cite{destexhe2009wilson}.

\subsubsection{Dynamical causal models}
\label{subsubsection:DcM}
Deducing the effective connectivity of functionally-segregated brain regions is crucial in developing bio-plausible and explainable models. Dynamical Causal Modeling (DCM), introduced in \cite{FRISTON2003DCM},  quantitatively generates the connectivities that fit the observed data by maximizing model evidence, aka marginal likelihood of the model \cite{Daunizeau2011DCM}. 

In a graph where the nodes are functionally-segregated populations, the effective connectivities are found based on three sets of parameters: (1) anatomical and functional couplings, (2) induced effect of stimuli, and (3) the parameters that describe the influence of the input on the intrinsic couplings. Models of the intra-connected regions can be built based on the earlier subsections, e.g., neural mass model, neural fields, or conductance-based models. For a review of such hybrid approaches, see \cite{Moran2013neural}.

\NewTColorBox{NewBox1}{ s O{!htbp} }{%
  floatplacement={#2},
  IfBooleanTF={#1}{float*,width=\textwidth}{float},
  colframe=cyan!75!black,colback=cyan!5!white,% any tcolorbox options here
  }

\begin{NewBox1}*
\begin{center}
    {\textbf{Neuromorphic compute: Architectures tailored for spiking networks}}
% Spiking neural networks, in abstract form or as giant hyper-realistic projects, are run by intensive parallel computing.
% In fact, many of the whole-brain simulations take place on supercomputers. However, beyond increasing the compute power and dataset size, attention to the medium of the simulations is necessary for achieving in-silico cognition.  

The disparity in energy consumption and computing architecture of biological and silicon neurons are the most important factors that raise eyebrows in assessing \textit{brain-like} algorithms.
The brain consumes $\sim$20 watts of power while this amount for a supercomputer is in order of megawatts \cite{zhang2018survey}. This twist verifies that the processing of information in these simulations is far from the biological truth. Apart from the energy consumption gap, the non-Von Neumann architecture of the brain is another discrepancy that stands in the way of realistic brain simulation in silico. There is no \textit{Von Neumann bottleneck} in the brain as there is no limitation on throughput as a result of separation of memory and computing unit \cite{wulf1995hitting}.
The brain also has other features that are greatly missed in deep networks. These include synaptic plasticity, high parallelism due to a large number of neurons, high connectivity due to a large number of synapses,  resilience to degradation, and low speed and frequency of communication, among other things. Although many of the aspects of biological cognition are complicated to reconstruct (e.g., embodiment and social interaction), the research in neuromorphic computing is addressing the disparities above by targeting hardware design \cite{cai2021neuromorphic}.

%Computation-intensive methods also pose difficulties in linking Spatio- and spectro-temporal brain data (STBD) to stimuli and output. The problem is that not only do they ignore almost all the biological assumptions, but a specific computational framework is also needed to deal with large-sized STBDs. 
A potential solution for narrowing this computation gap can be sought at the hardware level. An instance of such a dedicated pipeline is neuromorphic processing units (NPU) that are power efficient and take time and dynamics into the equation from the beginning. An NPU is an array of neurosynaptic cores that contain computing models (neurons) and memory within the same unit.
In short, the advantage of using NPUs is that they resemble the brain more realistically than a CPU or GPU because of asynchronous (event-based) communication, extreme parallelism (100-1000000 cores), and low power consumption \cite{Eliasmith2013How}. Their efficiency and robustness also result from the Physical proximity of the computing unit and memory. 
Below popular examples of such NPUs are listed. Each of them stemmed from different initiatives.
\begin{itemize}
    \item \textbf{SpiNNaker} or ``Spiking Neural Network architecture'' is an architecture based on low-power microprocessors and was first introduced in 2005 to help the European Brain Project with computations of large cortical area. The first version could imitate ten thousand spiking neurons and four million synapses with  43 nano Joules of energy per \textit{synaptic event}. \cite{sharp2012power}.

    \item \textbf{TrueNorth} chips are arrays of 4096 neurosynaptic cores amounting to 1 million digital neurons and 256 million synapses. IBM builds TrueNorth primarily as a low-power processor suitable for drones, but it is highly scalable and customizable \cite{akopyan2015truenorth}.  
    \item \textbf{Loihi} chips have demonstrated significant performance in optimization problems. Intel's fifth NPUs has incorporated biophysical reconstruction of hierarchical connectivity, dendritic compartments, synaptic delays, reward traces. Its circuit is composed of \textit{dandrite units} (for updating state variables), \textit{axon units}  (generating feed for the subsequent cores), and \textit{learning unit} (for updating weights based on customized learning rules)
    \cite{davies2018loihi}

    % \item \textbf{Neurogrid}
    % \item \textbf{Tianjic} 
\end{itemize}
% NPUs have Spiking information processes and are accessible through packages like Nengo \cite{Eliasmith2003NEF}.
% Nengo is a simulator that leverages biophysical priors \cite{Eliasmith2003NEF} (given in \cite{Eliasmith2014nengo} of single neurons to build a large-scale modeling framework
% they open the possibility of working with a variety of networks. The largest brain modeled so far has happened on NPUs with 6.6 million neurons, 20 billion connections, and 12 tasks.
An integrative example of the implementation discussed above is  NeuCube.
NueCube is a 3D SNN with plasticity that learns the connections among populations from various STBD modulations such as EEG, fMRI, genetic, DTI, MEG, and NIRS. Gene regulatory networks can be incorporated as well if available. Finally, This implementation reproduces trajectories of neural activity. 
It has more robustness to noise and higher accuracy in classifying STBD than standard machine learning methods such as SVM \cite{kasabov2014neucube}. 

Beyond biological alikeness, neuromorphic computing has important technical aspects that are missing in conventional compute units and can revolutionize neural data processing. They demonstrate lower latency, power consumption, and high portability required for real-time interpretation. These attributes make them useful for recent signal collectors like wearable EEG. On the other hand, although they have shown to be highly scalable and adaptable, their high cost per bit is a major pitfall \cite{davies2021lessons, sharifshazileh2021electronic}.  

\end{center}
\end{NewBox1}

\subsubsection{Spiking neural network: Artificial neural networks as a model of natural nervous system}

% With introduction of perceptron \cite{Rosenblatt1958perceptron} 
With the introduction of neural networks, the idea of implementing neural circuits and biological constraints into the artificial neural networks (ANN) gained momentum. \cite{McCulloch1943logical} is an early example that uses ANN with threshold spiking behavior. Despite being oversimplified, their idea formed the basis for a particular type of trainable network known as spiking neural networks (SNN) or biological neural networks (as in \cite{vreeken2003spiking}). Note that the distinction here with the other forms of spiking networks like Izhinevich's and derivatives (discussed earlier in Subsection \ref{subsection:Population_bio}) is that here we are talking about the networks that demonstrate a function approximation as a deep learning algorithm would do.

In contrast to deep neural networks, the activity in this architecture (transmission) is not continuous in time (i.e., during each propagation cycle). Instead, the activities are event-based occurrences with the event being the action potential depolarization\footnote{For a more comprehensive overview on types and applications, see \citet{Schliebs2013Evolving}.}. Although ANN architectures that are driven by spiking dynamics have been long used for optimization problems such as pattern recognition \cite{kasabov2007evolving} and classification \cite{soltic2008evolving}, they lag behind conventional learning algorithms in many tasks, but that is not the end of the story.

\citet{maass1997networks}  argues that concerning network size, spiking networks are more efficient in computation compared to other types of neural networks such as sigmoidal. Therefore it is worthwhile to implement SNNs in a more agnostic manner as spiking RNNs. Examples of such promising implementations are reservoir computing, liquid, and each state machine. For more on such neuromorphic architectures, see \ref{subsubsec:RNN}.
% In the previous section, we reviewed some of such realistic simulations (see \cite{deGaris2010LargeScaleSurvey} for more). The aforementioned models do not focus on linking behavior to the mechanism. What follows is a review of higher-order cognitive reconstructions based on spiking neuronal activities.

\subsection{Brain Atlases: whole- and population-level modeling}
The $21^{st}$ century has been the bursting era of large-scale brain initiatives. The objective of the simulation partly justifies this multitude. As it was previously mentioned, the notion \textit{simulation} is highly versatile in meaning depending on the goal of the project \cite{deGaris2010LargeScaleSurvey}, i.e., where it sits on the Figure \ref{fig:scales_levels}. Some of the projects of this spectrum are listed below.

\begin{itemize}
    \item BigBrain: a free-access and  few-cell-resolution model of human brain in 3D \cite{Landhuis2017Big}.
    \item Allen Brain Atlas: genome-wide map of gene expression for human adult and mouse brain \cite{jones2009allen}.
    \item Human Connectome Project: a large-scale structural and functional connectivity map of human brain [coined as connectome in \cite{Sporns2005Connectome}] \cite{VanEssen2013HCP}.
    \item Brain Research through Advancing Innovate Neurotechnologies: BRAIN \cite{Devor2013BRAIN_challenge}.
    \item The Virtual Brain (TVB): an open-source neural dynamics simulator using real anatomical connectivity \cite{jirsa2010towards}.
    \item Human Brain Project (HBP): aimed to realistically simulate the human brain in supercomputers \cite{Miller2011HBP_response}.
\end{itemize}
% BigBrain (a free-access few-cell-resolution three-dimensional (3D) model of human brain) \cite{Landhuis2017Big}, Allen Brain Atlas (genome-wide map of gene expression for human adult and mouse brain) \cite{jones2009allen}, Human Connectome Project (a large-scale structural and functional connectivity map of human brain [coined as connectome in \cite{Sporns2005Connectome}] \cite{VanEssen2013HCP}, Brain Research through Advancing Innovate Neurotechnologies (BRAIN) \cite{Devor2013BRAIN_challenge}, the Virtual Brain (an open-source GUI of large-scale connectivity matrices of realistic neural population for deriving dynamics) \cite{jirsa2010towards}, and perhaps most ambitious one, Human Brain project (aimed to realistically simulate human brain in super computers) \cite{Miller2011HBP_response}.

%In contrast to early studies (e.g., \cite{jirsa1996field, jirsa1997derivation, nunez1974brain}),
 The computing power is not the only problem when expanding the regional computations to the whole-brain level, scaling. One new difficulty is the integration of time delays that become significant at the whole-brain level. In local connections, the time delays are small enough to be ignored \cite{jirsa2010towards} the transmission happens in a variety of finite speeds from 1 to 10 meters 
per second. As a result of this variation, time delays among different brain parts are no longer negligible. Additional spatial features emerge by the implementation of this heterogeneity \cite{jirsa2000spatiotemporal, Petkoski2019Transmission}.

Larger scale approaches could adapt neural mechanisms that rely on intra-region interactions \cite{da1991neural} in order to ditch the problems related to the synaptic level studies mentioned earlier. The Virtual Brain (TVB) project is one of these initiatives. TVB captures the network dynamics of the brain by stimulating the neural population structural connectivity, the variant time scales, and noise \cite{sanz2013virtual}. 
TVB allows testing subject-specific hypotheses as the structural connectivity is based on individual DTI.
The large-scale activity is an integration of local neural masses connected through large-range dynamics.  
It has a web platform GUI and can run on a personal computer and has already implemented many types of dynamics for different types of brain signals, namely EEG, MEG, BOLD, fMRI.

With models like TVB, one should note the shift in paradigm from the fine-scale simulations like Blue Brain. 
Contrary to the Blue Brain, the nodes consist of large groups of neurons (order of a few millimeters), not one or a few neurons. Consequently, the governing equations are the ones for deriving population dynamics and statistical modes.
Another essential point is that TVB allows researchers to study the brain's phenomenology parametrically. The following section is dedicated to such studies.

\section{Phenomenological Models}
\label{sec:phenomenological}

% \subsection{Definition}
In contrast with realistic biological models of \textit{in-vivo} events, phenomenological\footnote{Note that here, the notion of "phenomena" here is different than that used by e.g. \cite{revonsuo2006inner} where phenomenological architecture and properties are regarded as a representation of environment in the first-person mind \cite{Stanford2018phenomenology}, complementary to ``physiological'' architecture in the brain–as in \cite{Fingelkurts2009Phenomenological}.} models offer a way of qualitative simulation of certain observable behavior (or, as it is discussed in Dynamical Systems literature \cite{strogatz2018nonlinear}, phase trajectories). The key assumption is that although short- and long-range dynamics are dependent on intricate biophysical events, the emerging observables can be encoded in significantly lower dimensions. This dimensionality reduction is thanks to dynamics that are capable of constructing similar statistical features of interest. Since a detailed-enough biophysical model should eventually exhibit the same collective statistics, one may argue that the phenomenological models offer a detour to system-level reconstruction by ditching lots of cellular and physiological considerations.

Compared to detailed biophysical models, coarse-grained approaches rely on a smaller set of biological constraints and might be considered ``too simplistic''. However, they are capable of reconstructing many collective phenomena that are still inaccessible to hyper-realistic simulations of neurons \cite{piccinini2021noise}. A famous example of emergence at this level is synchronizations in cortex \cite{Arenas2008Synchronization}. Moreover, experiments show that the population-level dynamics that are ignorant about the fine-grained detail \textit{better} explain the behavior  \cite{briggman2005optical, churchland2012neural}.

The significance of phenomenological models in the reconstruction of brain dynamics is also because of their intuitiveness and reproducibility. They may demonstrate critical properties of the neuronal population. An interesting example is noise-driven dynamics of the brain, which is responsible for multistability and criticality in resting-state \cite{deco2012ongoing, deco2017dynamics}. 
% It is fair to say that while predicting brain data as a many-body complex system is impossible, the qualitative reproduction of dynamics is an insightful way of inferring and controlling high-level cognition.

\subsection{Problem formulation, data, and tools}

The idea of using phenomenological models for neural dynamics is mainly motivated by the possibility of using tools from dynamical system theory. The goal is to quantify the evolution of a \textit{state space} built upon the \textit{state variables} of the system. 
For example, if one can find two population variables $(x,y)$ that determine the state of a neural ensemble, then all the possible pairs of $x$ and $y$s form the basis for the \textit{state space} of the system, let us call this 2-dimensional space $A$. The state of this ensemble at any given time $t$ can always be expressed as a 2-D vector in $A$. In mathematics, $A$ is called a \textit{vector space} defined by the sets of differential equations that describe the evolution of $x$ and $y$ in time. As an intuitive visualization of a vector space, imagine a water swirl: each point of the surface of a water swirl can be represented by a vector with the magnitude and direction of local velocity. One can see how at each point in this space, there is a \textit{flow} that pushes the system in a specific trajectory. 
% Because of the existence of chaos and other forms of nonlinearity \cite{izhikevich2000neural} the final state is usually inaccessible
Reproducing features of the brain signals or identifying such a sparse state space and the dynamics of a parsimonious set of state variables allows for forecasting the fate of the neural ensemble in future timesteps \cite{saggio2020phenomenological}. The evolution of the state variables is described by differential equations.
In what follows in this section, some of the most prominent phenomenological models and their findings are discussed.

Distinct dynamics is observed in a wide set of settings from resting-state activity \cite{piccinini2021noise} to task-specific experiments \cite{Pillai017Symmetry}. These insights are also useful in a wide set of modularities including fMRI \cite{Hutchison2013Dynamic}, EEG \cite{Atasoy2018Harmonic}, MEG \cite{Tait2021Systematic}, and Calcium imaging \cite{Abrevaya2021Learning}.
One common formulation is to build the dynamical graph models of the cortex based on the \textit{anatomical}, \textit{functional}, or \textit{effective connectivity} as described in Table \ref{table:Complex_brain_networks}. For a more comprehensive review of such networks, refer to \cite{Wein2021Brain}.

Connectivity matrices introduced in Table \ref{table:Complex_brain_networks} are the backbone of the information process pipeline. That being said, this parameter needs to be married to the dynamics of the states in the brain. To date, a large portion of studies have focused on mapping these networks onto the resting-state network, and a lot of structure-function questions remained to be answered by studying the task-related data \cite{Cabral2017Function}. In what follows, the models that quantify these dynamics based on the phenomenology of the behavior are discussed.

\begin{table*}[t]
    \centering
    \begin{tabular}{|c| c| c| c|} 
        \hline
        Connectivity & Measure & Imaging basis & Features \\ [0.5ex] 
        \hline\hline
        SC & \makecell{Spatial config. of \\white matter fibers} & \makecell{Static spatial images\\(e.g. DTI)} & Provides the anatomical architecture \\ 
        \hline
        FC & \makecell{Temporal correlations of\\regional activities} &  \makecell{Spatio-temporal images \\(e.g. fMRI, EEG) }& \makecell{Can be static or dynamic \cite{Surampudi2019Resting}\\Prone to spurious connections}\\
        \hline
        EC & \makecell{Causal interactions of\\segregated regions} & \makecell{Spatio-temporal images \\with a generative model \\(e.g. Granger Causality  \cite{ding200617}\\or DCM [\ref{subsubsection:DcM}])} & Rules out non-causal correlations  \\
        \hline
    \end{tabular}
    \caption{\textit{Complex brain networks are measured through Structural Connectivity (SC), Functional Connectivity (FC), or Effective Connectivity (EC). Computational Connectomics is a common ways of formulating structural and functional networks of the whole brain. Together with theories of dynamical graphs, these representations can provide insights into the collective faith of the system.}}  
    \label{table:Complex_brain_networks}
\end{table*}

\subsubsection{Generative graph models}
Recent progress in the science of complex networks and information theory has paved the way for analytical and numerical models of structural and functional connectivity \cite{Lurie2020questions}. The network approach to the neural population is a conventional way to study neural processes as information transmission in time-varying networks. This analogy allows for examining the path and behavior of the system in terms of different dynamical properties.

An insightful interplay of function versus structure is observed along the biologically-plausible line of works of Deco et al. \cite{deco2012ongoing}. They reconstructed the emergence of equilibrium states around multistable attractors and characteristic critical behavior like scaling-law distribution of inter-brain pair correlations as a function of global coupling parameters. 
% by studying multistability and synchrony \cite{Arenas2008Synchronization}. 
Furthermore, new studies show that synchrony not only depends on the topology of the graph but also on its hysteresis \cite{Qian2020Path}.

% and control \cite{liu2011controllability}. 
% The anatomical structure can be constructed from white matter streamlines tracked in diffusion tensor imaging \cite{honey2009predicting} in various granularity of nodes: from neurons to brain regions.
Tools from graph theory and network science \cite{newman2006structure} are used to formulate this relation. Spectral mapping \cite{becker2018spectral} and structure-function topological mapping \cite{liang2017structure} are proofs of concept in this regard.
Generative graph models (traditionally developed by graph theory such as the one for random graph introduced in \cite{erdHos1960evolution}) are principle tools of inference in this approach and now has been enhanced by machine learning, see for example, deep-network generative models in \cite{kolda2014scalable, li2018learning}). Simulations of brain network dynamics and study of controllability 
% (ease of switching dynamics)
\cite{kailath1980linear} has shown how differently regions are optimized for diverse behavior \cite{Tang2017developmental}.  
 
\subsection{Inspirations from statistical physics and nonlinear dynamical models}
In addition to network science, another axis for interpreting neural data is based on well-established tools initially developed for parametrizing the time evolution of physical systems. Famous examples of these systems include spin-glass  \cite{deco2012anatomy}, different types of coupled oscillators \cite{cabral2014exploring, Abrevaya2021Learning}, and multistable and chaotic many-body systems \cite{piccinini2021noise,  deco2017dynamics}. This type of modeling has already offered promising and intuitive results. In the following subsections, we review some of the recent literature with various methodologies.

\subsubsection{Brain as a complex system}
It is not easy to define what a complex system is. \cite{Haken2006Information} defines the \textit{degree of complexity} of a sequence as the minimum length of the program and of the initial data that a Turing machine (aka universal computer) needs to produce that sequence. Despite being a debatable definition, one can conclude that according to it, the spatiotemporal dynamics of the mammalian brain qualifies as a complex system \cite{sforazzini2014distributed, hutchison2011resting}. Therefore, one needs a complex mechanism to reconstruct neural dynamics. In the following few subsections, we review candidate equations for the oscillations in cortical network \cite{buzsaki2004neuronal}.
\paragraph{Equilibrium solutions and deterministic chaos}
Whole-brain phenomenological models like the Virtual Brain \cite{sanz2013virtual} are conventional generators for reconstructing spontaneous brain activity. There are various considerations to have in mind to choose the right model for the right task. A major trade-off is between the complexity and abstractiveness of the parameters \cite{breakspear2017dynamic}. In other words, to capture the behavior of detailed cytoarchitectural and physiological make-up with a reasonably-parametrized model. Another consideration is the incorporation of noise which is a requirement for multistable behavior \cite{piccinini2021noise} i.e., transitions between stable patterns of reverberating activity (aka attractors) in neural population in response to perturbation \cite{Kelso2012multistability}.
% https://www.nature.com/articles/s41598-017-03073-5 and the opened tabs

\paragraph{Kuramoto}
Kuramoto model is a mathematical descriptor of coupled oscillators, one that can be written down as simple as a system of ODEs solely based on sinusoidal interactions \cite{kuramoto1984chemical, nakagawa1994collective}.
Kuramoto model is widely used in physics for studying synchronization phenomena. It is relevant to neurobiological systems as it enables a \textit{phase reduction approach}: Neural populations can be regarded as similar oscillators that are weakly coupled together.
These couplings are parameterized in the model. Kuramoto can be extended to incorporate anatomical and effective connectivity and can expand from a low-level model of few-neuron activity to a stochastic population-level model with partial synchrony and rich dynamical properties. One way to do that is to upgrade the classic linear statistics to nonlinear Fokker-Planck equations \cite{Breakspear2010generative}.

There is significant literature on Kuramoto models on neural dynamics on different scales and levels. \cite{STROGATZ20001Kuramoto} is a conceptual review of decades of research on the principles of the general form of the Kuramoto model. Numerous studies have found consistency between the results from Kuramoto and other classic models in computational neuroscience like Wilson-Cowan \cite{wilson1972excitatory} \cite{HOPPENSTEADT199885thalamo}. Kuramoto model is frequently used for quantifying phase synchrony and for controlling unwanted phase transitions in neurological diseases like epileptic seizures and Parkinson's \cite{BOARETTO201protocol}. Still, there are many multistability questions regarding cognitive maladaptation yet to be explored, potentially with the help of Kuramoto models and the maps of effective connectivity. \cite{anyaeji2021quantitative} is a review targeting clinical researchers and psychiatrists. It is a good read for learning about the current challenges that could be formulated as a Kuramoto model.
Kuramoto is also unique in adaptability to different scales: from membrane resolution with each neuron acting as a delayed oscillator \cite{Hansel1995Synchrony} to the social setting where each subject couples with the other one in the dyad by means of interpersonal interactions \cite{dumas2012anatomical}.

\paragraph{Van der Pol}
Another model relevant to neuroscience is the van der Pol (VDP) oscillator which is probably the simplest relaxation oscillator~\cite{guckenheimer2013nonlinear} and a special case of the FitzHugh-Nagumo model, which is, in turn, a simplification of the
Hodgkin–Huxley model (see Subsection \ref{subsection:bio_history}) \cite{FITZHUGH1961impulses}. Through the Wilson-Cowan approximation \cite{kawahara1980}, VDP  can also model neural populations. For more information about the Wilson-Cowan model, please see Subsection \ref{subsubsection:Wilson-Cowan}. Recently, \cite{Abrevaya2021Learning} have used coupled VDP oscillators to model a low-dimensional representation of neural activity in different living organisms (larval zebrafish, rats, and humans) measured by different brain imaging modalities, such as calcium imaging (CaI) and fMRI. Besides proposing a method for inferring functional connectivity by using the coupling matrices of the fitted models, it was demonstrated that dynamical systems models could be a valuable resource of data augmentation for spatiotemporal deep learning problems.

Looking at the brain as a complex system of interacting oscillators is a detour for expanding the modeling to larger organization scales.
The emergent behavior of the system can be described with ``order parameters''. Although this is a description with much lower dimensions than the biophysical equations, it still expresses many remarkable phenomena such as phase transitions, instabilities, multiple stable points, metastability, and saddle points \cite{Haken2006Information}. However, parametrizing such models is still an ongoing challenge, and many related studies are limited to the resting-state network. The following section reviews the prospects of recent data-driven methods and how they can leverage the study of system-level behavior.

% \paragraph{Van der Pol}
% Van der Pol (VDP) model, one of the simplest relaxation oscillators~\cite{guckenheimer2013nonlinear}, is relevant to neuroscience as a special case of the FitzHugh-Nagumo model, which is, in turn, a simplification of the classical
% Hodgkin–Huxley model of activation and deactivation dynamics of a spiking neuron (see \ref{subsection:bio_history}) \cite{FITZHUGH1961impulses}.  VDP  can also model neural populations through its approximation of Wilson-Cowan dynamics \cite{kawahara1980}. For more information about the Wilson-Cowan model, please see \ref{subsubsection:Wilson-Cowan}. Recently, \cite{Abrevaya2021Learning} have used coupled VDP oscillators to model a low-dimensional representation of neural activity measured by different brain imaging modalities, such as calcium imaging (CaI) and fMRI, in different living organisms: larval zebrafish, rats, and humans. Besides proposing a method for inferring functional connectivity by using the coupling matrices of the fitted models, it was demonstrated that dynamical systems models could be a valuable resource of data augmentation for spatiotemporal deep learning problems.

% \subsection{Ising}

\section{Agnostic Computational Models}
    \label{sec:agnostic_computational}
    
Jim Garys's framework \cite{hey2009fourth} divides the history of science into four paradigms. Since centuries ago, there have been experimental and theoretical paradigms. Then the phenomena of interest became too complicated to be quantified analytically, so the computational paradigms started with the rise of numerical estimations and simulations.
Today, with the bursting advances in recording, storage, and computation capacity of neural signals, neuroscience is now exploring the \textit{fourth paradigm} of Jim Garys's framework \cite{hey2009fourth} i.e., data exploration in which the scientific models are fit to the data by learning algorithms.

In the introduction\ref{sec:Intro}, we reflected on how scientists should not settle for mere prediction.
While the literature on data-driven methods is enormous, this review focuses mainly on the strategies that help gain mechanistic insights rather than those that reproduce data through operations that are difficult to relate to biological knowledge.
Instances of these unfavored methods include strict generative adversarial networks with uninterpretable latent spaces or black-box RNN with hard-to-explain parameters.
The following section categorizes these methods into established and emerging techniques and discusses some showcases.

%%%%%%%%%%%%%%%%%%%%%%%%%%%%%%%%%%%%%%%%%
% TODO: This is a bunch of "good to know" stuff that I can't really fit into the flow:
% The notion of the practicability of a model is highly goal-specific.
% Neural populations are self-organizing systems that continuously interact with the environment. Therefore, any signal is a result of intrinsic dynamics, externally-evoked stimuli, or both.
% This distinction is crucial in choosing the methodology.

% Another factor to contemplate is the assessment of models. In other words, to define \textit{better} and \textit{worse} in the context of modeling and to guide the optimization process. 
% Mean square error cannot be deployed as an empirical measure for complex systems with chaotic dynamics, mainly because of the accuracy limit for initial conditions and parameters. Kullback-Leibler divergence \cite{koppe_identifying_2019}, $l_2$ error in eigenvalue estimation \cite{hsu2020linear}, representational similarity Analysis (mostly for representational models) \cite{kriegeskorte2008representational}, and supervised classifiers such as Multi-Voxel Pattern Analysis \cite{davatzikos2005classifying, norman2006beyond} are proposed as measures that capture ‘ground-truth dynamical properties. 
%%%%%%%%%%%%%%%%%%%%%%%%%%%%%%%%%%%%%%%%%

% \subsection{Task-based Inference}
\subsection{Established learning models}
% These models are particularly (but not exclusively) useful for cognitive, motor, and visual cortex, regenerating representations/patterns. Such dynamical generative models
% are mostly based on Deep Learning (DL) methods because of some reasonable (but sometimes distant) similarities of information flow in deep nets and natural neural nets.

Data-driven models have long been used in identifying structure-function relations (similar to the ones mentioned in Table \ref{table:Complex_brain_networks}) \cite{mckeown1998analysis, koppe_identifying_2019}. The shift of studies from single-neuron to neural network, or more precisely, has accelerated in the last decade. This trend is because relying on collective properties of a population of neurons to infer behavior seems more promising than reconstructing the physiological activity of single neurons in hopes of achieving emergence. Yuste \cite{yuste2015neuron} argues that the mere representations that relate the state of individual neurons to a higher level of activity have serious shortcomings \cite{michaels2016neural}. However, these shortcomings can be addressed by incorporating temporal dynamics and collective measures into the model. We review the models that satisfy this consideration.

\subsubsection{Dimensionality Reduction Techniques} 
% From https://www.hindawi.com/journals/cin/2021/5573740/
Clustering and unsupervised learning are useful for mapping inputs ($\boldsymbol{X}$) to features ($\boldsymbol{Y}$). Later, this set of $(\boldsymbol{X},\boldsymbol{Y})$ can be extrapolated to unseen data.  
There are various methods for identifying this mapping or, in other words, for approximating this function.
Principal Component Analysis (PCA) is a primary one. % TODO Citation for PCA?
PCA maps data onto a subspace with the maximal variance \cite{Markopoulos2014optimal}. It is a common method of dimensionality reduction. However, the orthogonal set of features found by this method are not necessarily statistically-dependent. Therefore, they are not always helpful in finding sources and effective connectivity. Alternatively, Independent Component Analysis, commonly known as ICA, was introduced as a solution to the Blind Source Separation (BSS) problem. Each sample of the data is an ensemble of the state of different sources. However, the characteristics of these sources are the hidden variable \cite{pearlmutter1997maximum}. ICA is effective in finding the related source as it maps the data onto the feature space by minimizing the statistical independence for each feature rather than by minimizing the variance.
Conventional use of component analysis is with fMRI and EEG recordings. In each time window, each sensor receives a noisy mix of activity in segregated brain regions. One is usually interested in inferring effective connectivity based on such data. Having a large number of electrodes around the scalp enables ICA to identify the independent sources of activities and artifacts. ICA algorithms come in different flavors depending on the dataset and the property of interest. E.g. temporal- \cite{calhoun2001spatial}, spatial- \cite{mckeown1998analysis}, and spatiotemporal-ICA \cite{goldhacker2017multi,wang2014temporally} are tailored for different types of sampling. Hybrid approaches, e.g., ICA amalgamated with structural equation modeling (SEM), have shown better performance in given setups with less prior knowledge than SEM alone \cite{rajapakse2006exploratory}. 
% Component analysis method can also help decompose the functional connectivity evolutions into hidden state-space models, e.g., with the help of Markov chain processes as in \cite{eavani2013unsupervised}.
The interested reader is encouraged to refer to \cite{calhoun2012multisubject} for a dedicated review of ICA methods.

\subsubsection{Recurrent neural networks}
\label{subsubsec:RNN}
% Many of the sequence models in computational neuroscience stem from the ones initially developed for natural language processing. For example, algorithms like Word2Vec \cite{mikolov2013distributed} that were exclusively developed to capture semantic links and relations in the text have been modified by \citet{rosenthal2018mapping} to embed functional connectivity or connectome in a significantly-low dimensional vector space to predict functional and structural topology.

Recurrent neural networks (RNN) are the Turing-complete \cite{Kilian1996Dynamic} algorithms for learning dynamics and are widely used in computational neuroscience.
In a nutshell, RNN processes data by updating a ``state vector''. The state vector holds the memory across steps in the sequence. This state vector contains long-term information of the sequence from the past steps \cite{LeCun2015deep}. 

Current studies validate diverse types of RNNs as promising candidates for generating neural dynamics. \citet{sherstinsky2020fundamentals} shows how the implicit ``{additive model}'', which evolves the state signal, incorporates some of the interesting bio-dynamical behavior such as saturation bounds and the effects of time delays. 
Several studies modeled the cerebellum as an RNN with granular \cite{buonomano1994neural, hofstoetter2002cerebellum, medina2000timing, yamazaki2005neural} or randomly-connected layers \cite{Yamazak2007Echo}.
Moreover, similarities of performance and adaptability to limited computational power (as in biological systems) are observed both in recurrent convolutional neural networks and in the human visual cortex \cite{spoerer2020recurrent}.

RNNs vary greatly in architecture. The choice of architecture can be implied by the output of interest (for example text \cite{sutskever2011generating} versus natural scenes\cite{socher2011parsing})
or the approaches to overcome the problem with vanishing and exploding gradient (e.g., long short-term memory (LSTM) \cite{hochreiter1997long}, hierarchical \cite{Hihi1996Hierarchical}, or gated RNNs \cite{chung2014empirical}). 

\textbf{Hopfield} Hopfield network \cite{Hopfield1982neural} is a type of RNN inspired by dynamics of Ising model \cite{little1974existence, brush1967history}. In the original Hopfield mechanism, the units are 
% binary 
threshold \citet{McCulloch1943logical} {neurons}, connected in a recurrent fashion. The state of the system is described by a vector $V$ which represents the states of all units. In other words, the network is in fact, an undirected graph of artificial neurons.
The strength of connection between units $i$ and $j$ is described by $w_{ij}$ which is trained by a given learning rule i.e. commonly Storkey \cite{storkey1997increasing} or {Hebbian rule} (stating that ``{neurons that fire together, wire together}'') \cite{hebb1949organisation}.
After the training, these weights are set, and an energy landscape is defined as a function of $V$. The system evolves to minimize the energy and moves toward the basin of the closest attractor. This landscape can exhibit the stability and function of the network \cite{yan2013nonequilibrium}.

The Hopfield model can accommodate some biological assumptions and work in tandem with cortical realizations. Similar to the human brain, Hopfield connections are mostly symmetric. Most importantly, since its appearance, it has been widely used for replicating associative memory. However, soon it was revealed that other dynamical phenomena like cortical oscillations and stochastic activity \cite{wang2010neurophysiological} need to be incorporated in order to capture a comprehensive image of the cognition.
% Modern Hopfield network or dense associative memory

\textbf{LSTM} 
In addition to the problem of vanishing and exploding gradient, other pitfalls also demand careful architecture adjustment. Early in the history of deep learning, RNNs demonstrated poor performance on sequences with long-term dependencies \cite{schmidhuber1992learning}. Long short term memory (LSTM) is specifically designed to resolve this problem. The principle difference of LSTM and vanilla RNN is that instead of a single recurrent layer, it has a ``{cell}'' composed of four layers that interact with each other through three gates: input gate, output gate and forget gate. These gates control the flow of old and new information in the ``{cell state}'' \cite{hochreiter1997long}. On certain scales of computation, LSTM still has considerable performance compared to trendy sequential models like transformers.

\textbf{Reservoir Computing}
A reservoir computer (RC) \cite{maass2002real} is an RNN with a reservoir of interconnected spiking neurons. Broadly speaking, the distinction of RC among RNNs, in general, is the absence of granular layers between input and output.
RCs themselves are dynamical systems that help learn the dynamics of data.
Traditionally, the units of a reservoir have nonlinear activation functions that allow them to be universal approximators. \citet{Gauthier2021next} show that this nonlinearity can be consolidated in an equivalent \textit{nonlinear vector autoregressor}. With the nonlinear activation function out of the picture, the required computation, data, and metaparameter optimization complexity are significantly reduced, interpretability is consequently improved while the performance stays the same.

\textbf{Liquid state machine} LSM can be thought of as an \textit{RNN soup} that maps the input data to a higher dimension that more explicitly represents the features. The word \textit{liquid} come from the analogy of a stone (here an input) dropping into the water (here a spiking network) and propagating waves. LSM maintains intrinsic memory and can be simplified so much that it processes real-time data \cite{Polepalli2016digital}. 
\citet{Zoubi2018Anytime} shows LSM performs notably in building latent space of EEG data (extendable to fMRI). As for the faithfulness to the biological truth, Several studies argue that LSM surpass RNNs with granular layers in matching organization and circuitry of cerebellum \cite{Yamazak2007Echo} and cerebral cortex \cite{maass2002real}. \citet{lechner2019designing} demonstrate the superiority of a biologically-designed LTM on given accuracy benchmarks to other ANNs, including LSTM.  

\textbf{Physically-informed RNN} A prominent factor in determining the dynamical profile of the brain is the intrinsic time delays \cite{chang2018reversible}. Integrating these time delays into the artificial networks was initially an inspiration from neuroscience for AI. Later, they came back as a successful tool for integrated sequence modeling for multiple populations. In the last decade, RNN has been used for reconstructing neural dynamics via interpretable latent space in different recording modalities such as fMRI \cite{koppe_identifying_2019} and Calcium imaging \cite{Abrevaya2021Learning}.
% integrating Spatio-temporal neural computation in various brain activities such as  
\\Continuity of time is another extension that can make RNNs more compatible with various forms of sampling and thus neural dynamics from spikes to oscillations. Continous time RNNs (CT-RNNs) are RNNs with activation functions made up of differential equations. They have been proved to be universal function approximators \cite{funahashi1993approximation} and have surfaced recently in the literature as {reservoir computers} \cite{verstraeten2007unifying, Gauthier2021next} and {liquid time-constant neural networks} \cite{hasani2020liquid}. 

\paragraph{}Essentially, finding the optimal architecture and hyperparameters for a given problem does not have a straightforward recipe. The loss function in a deep neural network can be arbitrarily complex and usually takes more than a convex optimization. \citet{Li2018Visualizing} shows how parameters of the network can change the loss landscape and trainability. Another more specific issue to the algorithms trained on a temporal sequence is catastrophic forgetting and attention bottleneck. These complications arise from the limitation of memory and attention to the past time steps. New attention models such as transformers and recurrent independent mechanisms (see Subsections \ref{subsubsec:transformers} and \ref{subsubsec:RIM} respectively) are specifically built to address these issues. As memory-enhanced components, RNN layers appear in other deep and shallow architectures with sequential data as input, including encoder-decoders.

\subsubsection{Variational autoencoder} 
% A famous method for 
    Variational autoencoders (VAE for short) is a type of neural network that encodes the ground truth as the input onto a ``latent space'' and then decodes that space for reconstructing the input\cite{kingma2013auto}. The network is parameterized by minimizing the reconstruction loss, which is, in this case, a metric of information gained by a metric called Kullback–Leibler divergence (\cite{kullback1951information}.
    An example of VAE used for regenerating dynamics is by \citet{perl2020generative} in which the coupling dynamics of the whole brain and the transitions between the states of wake-sleep progression is generated. The goal is to find low (e.g., as low as 2-) dimensional manifolds that can capture the signature structure-function relationship that demonstrates the stage in the wake-sleep cycle \cite{barttfeld2015signature, vincent2007intrinsic} and the parameters of
generic coupled Stuart-Landau oscillators as in \cite{deco2017dynamics}.
    % \\The long-range axon architecture (and consequently, the symmetric coupling matrix) is produced by Diffusion Tensor Imaging (DTI) recordings. Other parameters of the  Stuart-Landau equations -bifurcation parameter and global coupling scale- can be optimized by training on the correlation matrix of different populations based on fMRI signals. Similar to \cite{ipina2020modeling}, these neural populations are determined by Resting-State Network (RSN) functional divisions.
An idea for regenerating dynamics is to use a deep-network embedded differential equations (as in Subsection \ref{subsec:NDE}) in the latent VAE structure  \cite{chen2018neural}.

% TODO flesh this out
\subsubsection{Transformers}
\label{subsubsec:transformers}
The transformer is a relatively new class of ML models that recently has shown state-of-the-art performance on sequence modeling such as natural language processing (NLP) field of research \cite{vaswani2017attention}. Beyond NLP, this architecture demonstrates good performance on a wide variety of data, including brain imaging \cite{Jiayao2021EEGTransformer, song2021transformerbased, demetres2021bendr}. Similar to RNNs, transformers aim to process sequential data such as natural language or temporal signals. It differs from the RNN paradigm because it does not process the data sequentially; instead, it looks at whole sequences with a mechanism called ``attention,'' and by doing so, it alleviates the problem of forgetting long dependencies, which is common in RNN and LSTMs. This mechanism can make both long- and short-term connections between points in the sequence and prioritize them.
% This type of model made its mark modeling long sequences of data and generating sequences that are the most likely given the first part of the sequence.
Transformers are widely used for generating \textit{foundation models} (i.e. models that are pretrained on big data \cite{bommasani2021opportunities}) and they can outperform recurrent networks like LSTM with large models/data \cite{kaplan2020scaling}.

\subsubsection{Recurrent independent mechanisms}
\label{subsubsec:RIM}

Recurrent independent mechanisms (RIM) is a form of attention model that learns and combines independent mechanisms in order to boost generalisability and robustness in executing a \textit{task}. 
The task in the sense of signal processing can be generating a sequence based on the observed data. 
The hypothesis is that the dynamics can be learned as a sparse modular structure.
In this recurrent architecture, each module independently specializes in a particular mechanism. Then all the RIMs compete through an attention bottleneck so that only the most relevant mechanisms get activated to communicate sparsely with others to perform the task \cite{goyal2020recurrent}.

%%%%%%Defining SciML
\subsection{New frontiers: Scientific ML and Interpretable Models}

The independence from prior knowledge sounds interesting as it frees the methodology from inductive biases and makes the models more generalizable by definition. However, this virtue comes at the cost of a need for large training sets.
In other words, the trade-off of bias and computation should be considered: Applying lots of prior knowledge and inductive biases result in a lesser need for data and computation. In contrast, little to no inductive bias calls for a great need for big and curated data.
% NOT related(related to no-free lunch therem in machine learning first discussed in \cite{wolpert1996existence} and \cite{wolpert1996lack}.
It is true that with the advancement of recording techniques, the scarcity of data is less of a problem than it was before but even with all these advances, having \textit{clean} and \textit{sufficiently large} medical dataset that helps with the problem in hand is not guaranteed.

Total reliance on data is especially questionable when the data has significant complications (as discussed in the introduction\ref{sec:Intro}). Opting for a methodology guided by \textit{patterns} rather than \textit{prior knowledge} is problematic in particular when the principle patterns of data arise from uninteresting phenomena such as the particular way a given facility may print out the brain images \cite{ng2021AI}.

The thirst for data aside, one of the prominent drawbacks of agnostic modeling and ML, in particular, is that they are famous for providing \textit{opaque blackbox} solutions, meaning that by leaving biological priors out of the picture, the explainability of the outcome is weak. Lack of explainability is a pet peeve for people in science as they are interested in both prediction and the reasoning behind those predictions. 

% such as least absolute shrinkage and selection operator aka Lasso \cite{Tibshirani1996Regression}   
% problem $\boldsymbol{X} = \boldsymbol{AB}$

In addition to the implicit assumption of the adequacy of training data, the explicit assumption that these models rely on is that the solution is parsimonious, i.e., there are few descriptive parameters. Despite some possibility of error with this assumption in given problems \cite{su2017false}, it is particularly useful in having arbitrarily less complicated descriptions that are generalizable, interpretable, and less prone to over-fitting.

% Scientific machine learning amend the problem of interpretability and data-insensitivity by supplying the solver with background knowledge. For example,
% To tackle this, we argue that using generative models provides better 'understanding' as not only they can be regarded as predictive tools, but also because they allow for mimicking the system.

% \subsubsection{Jack of all trades, master of some: Data-driven models with minimal pre-assumptions}
% There are general function approximators that could identify data dynamics without injecting any prior knowledge about the system. Of course, the price is the need for large datasets. They could provide a perfect solution for a well-observed system with unknown dynamics. SINDy and neural ODEs are in this category. 

The following sections describe general function approximators that could identify data dynamics without injecting any prior knowledge about the system. They could provide a perfect solution for a well-observed system with unknown dynamics.

\subsubsection{Sparse identification of nonlinear dynamics}  

\citet{kaheman2020sindy} proposed a novel approach for quantifying underlying brain dynamics. The key assumption is that the governing multi-dimensional principles can be derived by a system of equations describing the first-order rate of change.
In order to use sparse regressions methods such as Sparse Identification of Nonlinear Dynamics (SINDy), one needs to precisely specify the set of parsimonious state variables \cite{quade2018sparse}.
That being said, SINDy does not work for small datasets. If it is given less data than possible terms, the system of the governing equations is under-specified. Therefore, the underfitting as a result of insufficient training data is the secondary problem. One approach to address this issue is incorporating the known terms and dismissing the learning for those parts. An example is discussed in Subsection \ref{subsubsec:UDE}.

\subsubsection{Differential equations with deep neural networks}
    \label{subsec:NDE}
    A relatively new class of dynamical frameworks combines differential equations with machine learning in a more explicit fashion. This class of frameworks has been used to model the dynamics of time-varying signals. They begin by assuming that the underlying dynamics follow a differential equation. They can then be used to discover the parameters of that differential equation by using standard optimization of deep neural networks. As is evident, such formulations are quite useful in modeling and analyzing brain dynamics, especially using deep networks. Below we describe some of the relevant works in this sub-field.
    \paragraph{Neural Ordinary Differential Equations}
Combining ordinary differential equations (ODEs) with neural networks has recently emerged as a feasible method of incorporating differentiable physics into machine learning. A Neural Ordinary Differential Equation (Neural ODE)~\cite{chen2018neural} provides the use of a parametric model as the differential function in an ODE. This architecture learns the dynamics of a process without explicitly stating the differential function, as has been done previously in different fields. Instead, standard deep learning optimization techniques could be used to train a parameterized differential function that can accurately describe the dynamics of a system. In the recent past, this has been used to infer the dynamics of various time-varying signals with practical applications~\cite{chen2018neural, nODEs_irregularly,kanaa2019simple,yildiz2019ode2vae,liu2020sde, li2020scalable, jia2019jump, kidger2020neural}.

    \paragraph{Latent ODE}
A dynamic model such as the Neural ODE can be incorporated in an encoder-decoder framework, resembling a Variational Auto-Encoder, as mentioned in~\citet{chen2018neural}. Such models assume that latent variables can capture the dynamics of the observed data. Previous works~\cite{chen2018neural, nODEs_irregularly,kanaa2019simple,yildiz2019ode2vae} have successfully used this framework to define and train a generative model on time series data.
    \paragraph{Stochastic neural ODEs}
Parametric models can also be incorporated into stochastic differential equations to make Neural Stochastic Differential Equations (Neural SDEs)~\cite{liu2020sde, li2020scalable}. Prior works have also introduced discontinuous jumps~\cite{jia2019jump} in the differential equations.

    \paragraph{Neural controlled differential equations}
Further, latent ODE models can add another layer of abstraction. The observed data is assumed to be regularly/irregularly sampled from a continuous stream of data, following dynamics described by a continuously changing hidden state. Both the dynamics of the hidden state and the relationship between the interpolated observations and the hidden state can be described by neural networks.
% and trained end-to-end.
Such systems are called Neural Control Differential Equations (Neural CDE) \cite{kidger2020neural}. Broadly speaking, they are the continuous equivalent of RNNs.

\subsubsection{Differential equations enhanced by deep neural networks}

The above methods use deep neural networks to define the differential function in ordinary differential equations. In contrast, UDEs and GOKU-nets (described below) take the help of deep neural networks to enhance differential equations. UDEs replace only the unknown parts of a known partial differential equation, while GOKU-nets use explicit differential equations as part of deep neural network pipelines.

\textbf{Universal differential equations:}
    \label{subsubsec:UDE}
Universal Differential Equations (UDE) offer an alternate way of incorporating neural networks into differential equations while accounting for prior knowledge. In their seminal work, \citet{Rackauckas2020Universal} demonstrate how it is possible to aid a partial differential equation model by learning the unknown terms with universal approximators such as neural networks. Furthermore, they show how by combining this approach with a symbolic regression, such as SINDy, these models can accelerate the discovery of dynamics in limited data with significant accuracy.
% Hence, UDEs take advantage of partially known governing laws of dynamical systems and fill the gaps with machine learning.

% the cost function for UDEs is defined as the Euclidean distance between the solution of the Differential Equation and the ground truth of the data point at the same time instant. 

% \subsubsection{Other Representational Learning approaches}
% %I.E. Tools in SciML, misc generative models (eg graph) etc

% \subsection{State and transition Inference}
% Strictly-dynamical models as needed in in resting-state. 
% Examples: looking at activity profile distributions across neural population and deriving representational geometry based on that \cite{kriegeskorte2019peeling}.
% These are useful for many purposes such as "architectures, learning rules and objective functions"
% TODO cite robustness (Lajoie's talk) and memory as examples.

% SINDy, SciML methods
% \subsection{classification}
% supervised (e.g. MVPA) or unsupervised (e.g. dimensionality reduction methods, SVM)
% "good-practice guidelines for unbiased application of generative embedding in the context of fMRI:" \cite{Brodersen2011}

% \begin{enumerate}
%     \item SVD
%     \item Multidimensional Scaling (MDS)
%     \item Manifold Learning
%     \item Diffusion Maps
% \end{enumerate}

\textbf{Generative ODE Modeling with Known Unknowns: }
Another promising approach is the case of the Generative ODE Modeling with Known Unknowns aka GOKU-nets \cite{GOKU}. GOKU-net consists of a variational autoencoder structure with ODEs inside. In contrast with Latent ODEs, here, the ODEs are not parameterized but given explicit forms. Hence, it is possible to make use of some prior knowledge of the dynamics governing the system, such as in SINDy and UDEs, but there is no need to have direct observations of the state variables as in those cases. For example, one could hypothesize that the latent dynamics of a system follow some particular differential model such as Kuramoto or van der Pol. This model then jointly learns the transformation from the data space to a lower-dimensional feature representation and the parameters of the explicit differential equation. 

\paragraph{}The machine learning techniques are now routinely used for classification and regression of brain states (see \citet{Wein2021Brain} for a review). However, they have much more potential than black-box, data-intensive classifiers. This is because new sequential models are sometimes designed to identify the missing pieces of the puzzle of dynamics. They can also act as generative models and provide a broad potential for testing the biophysical and system-level hypotheses.
Some of the methods introduced in this section are explained in detail in \citet{kutz2013data} textbook. Moreover, extremely helpful tutorials can be found in \citet{Brunton} YouTube\textsuperscript{TM} channel.

% The smallest interacting blocks of the nervous system are proteins \cite{Heuvel2019multiscale}. The genetic expression maps and atlases are useful for discovering the functions of these proteins in the neural circuit \cite{mazziotta2000probabilistic}.

\section{Conclusion}

The key purpose of this review was to dive into samples of already-popular paradigms or the ones authors found most promising for reconstructing neural dynamics with all the special considerations. To achieve this, we sorted the computational models with respect to two indicators: the scale of organization and the level of abstraction (Figure \ref{fig:scales_levels}).

The scope of our study is broadly generative models of neural dynamics in biophysics, complex systems, and AI with some limitations. This paper is an interdisciplinary study that covers a time span from the mid-twentieth century when the pioneer models like Wilson-Cowan \cite{wilson1972excitatory}, and Hodgkin-Huxley \cite{hodgkin1952quantitative} arose, up until the recent decade when gigantic brain atlas initiatives, groundbreaking research in ML, and unprecedented computation power became available. Given the rate of publication in the related fields, a systematic review was impossible. Therefore, this paper is a starting point for gaining an eagle-eye view of the current landscape. 
% There is no ``{silver bullet}'' and 
It is up to the reader to adjust the model scale and abstraction depending on the problem at hand (see Fig. \ref{fig:scales_levels}).

We emphasized the distinctiveness of the problems in computational neuroscience and cognitive science. One key factor is the trade-off of complexity and inductive bias with the availability of data and prior knowledge of the system. While there is still no ultimate recipe yet, hybrid methods could simultaneously tackle explainability, interpretability, plausibility, and generalizability. 

\section{Acknowledgments}
The authors are grateful for the discussions and revisions from Hadi Nekoei, Timothy Nest, Quentin Moreau, Eilif Muller, and Guillaume Lajoie. Guillaume Dumas is funded by the Institute for Data Valorization (IVADO), Montréal and the Fonds de recherche du Québec (FRQ). Mahta Ramezanian Panahi, Jean-Christophe Gagnon-Audet, Vikram Voleti, and Irina Rish acknowledges the support from Canada CIFAR AI Chair Program and from the Canada Excellence Research Chairs (CERC) program.

\section{Conflict of Interest Statement}
The authors declare that the research was conducted in the absence of any commercial or financial relationships that could be construed as a potential conflict of interest.
%----------------------
\bibliographystyle{unsrtnat}
\bibliography{survey_bib}
\end{multicols}
\end{document}